\documentclass[10pt,conference,a4paper,twocolumn]{IEEEtran}

\IEEEoverridecommandlockouts

\usepackage{cite}
\usepackage{amsmath,amssymb,amsfonts}
\usepackage{algorithmic}
\usepackage{graphicx}
\usepackage{textcomp}
\usepackage{hyperref}
\usepackage{booktabs}
\usepackage[ruled,linesnumbered]{algorithm2e}
\usepackage{xcolor}
\def\BibTeX{{\rm B\kern-.05em{\sc i\kern-.025em b}\kern-.08em
    T\kern-.1667em\lower.7ex\hbox{E}\kern-.125emX}}

\usepackage{xspace}
\makeatletter
\DeclareRobustCommand\onedot{\futurelet\@let@token\@onedot}
\def\@onedot{\ifx\@let@token.\else.\null\fi\xspace}

\def\eg{\emph{e.g}\onedot} 
\def\ie{\emph{i.e}\onedot}

\def\wrt{w.r.t\onedot} 
\def\etal{\emph{et al}\onedot}
\makeatother

\makeatletter
\def\ps@IEEEtitlepagestyle{%
  \def\@oddfoot{\mycopyrightnotice}%
}
\def\mycopyrightnotice{%
\begin{minipage}{\textwidth}
\centering \footnotesize
Copyright~\copyright~2023 IEEE. Personal use of this material is permitted.  Permission from IEEE must be obtained for all other uses, in any current or future media, including reprinting/republishing this material for advertising or promotional purposes, creating new collective works, for resale or redistribution to servers or lists, or reuse of any copyrighted component of this work in other works.
\end{minipage}
}
\makeatother

\newcommand{\pname}{\textit{HermesBDD}\xspace}
\newcommand{\ite}{\textsc{If-Then-Else}\xspace}
\newcommand{\nqueens}{$n$-Queens\xspace}

\begin{document}

\title{
HermesBDD: A Multi-Core and Multi-Platform Binary Decision Diagram Package
\thanks{
The work has been partially supported by the Italian Ministry of Education, University and Research (MIUR) with the grant ``Dipartimenti di Eccellenza'' 2018-2022, and by the project INdAM, GNCS 2020 (Strategic Reasoning and Automated Synthesis of Multi-Agent Systems) funded by MIUR (Italian Ministry of Education, University and Research).
}
}

\author{
\IEEEauthorblockN{Luigi Capogrosso, Luca Geretti, Marco Cristani, Franco Fummi, Tiziano Villa}
\IEEEauthorblockA{\textit{Department of Computer Science, University of Verona, Italy}}
{\tt name.surname@univr.it}
}

\maketitle
%Add space between copyright and text.
\IEEEpubidadjcol

%%%%%%%%% ABSTRACT.
\begin{abstract}
BDDs are representations of a Boolean expression in the form of a directed acyclic graph. BDDs are widely used in several fields, particularly in model checking and hardware verification. There are several implementations for BDD manipulation, where each package differs depending on the application. % The variation of package architectures comes from resizing criteria and realization and node manipulations.
This paper presents \pname{}: a novel multi-core and multi-platform binary decision diagram package focused on high performance and usability. \pname{} supports a static and dynamic memory management mechanism, the possibility to exploit lock-free hash tables, and a simple parallel implementation of the \ite{} procedure based on a higher-level wrapper for threads and futures. %
\pname{} is completely written in C++ with no need to rely on external libraries and is developed according to software engineering principles for reliability and easy maintenance over time. %
We provide experimental results on the \nqueens{} problem, the de-facto SAT solver benchmark for BDDs, demonstrating a significant speedup of $18.73\!\times{}\!$ over our non-parallel baselines, and a remarkable performance boost \wrt{} other state-of-the-art BDDs packages. %
% The source code is available at \url{https://luigicapogrosso.github.io/HermesBDD}.
\end{abstract}

%%%%%%%%% KEYWORDS.

%%%%%%%%% BODY TEXT.
\section{Introduction}
\label{cha:intro}

Binary decision diagrams (BDDs) were introduced by Akers~\cite{akers1978binary} and developed by Bryant~\cite{bryant1986graph} and provide a data structure for representing and manipulating Boolean functions. There are several implementations of BDD packages in the literature (details in Sec.~\ref{cha:related}), but they focus mainly on package performance regarding speedup and memory efficiency. However, we believe that performance is only one aspect to judge a BDD package. Other aspects, such as (not necessarily in order of importance), \textit{functionality}, \textit{robustness}, \textit{reliability}, \textit{portability}, and \textit{documentation}, matter as well. According to these principles, we developed \pname{}: a novel multi-core and multi-platform binary decision diagram package focused on high performance and usability. It supports a static and dynamic memory management mechanism, the possibility to exploit lock-free hash tables, and a parallel implementation of the \ite{} procedure based on a higher-level wrapper for threads and futures. Additionally, \pname{} presents a well-documented source code, it is completely written in C++ with no need to rely on external libraries, and it is developed according to engineering principles such as testability, code coverage, and continuous integration.

We provide experimental results on the \nqueens{} problem showing how our multi-core implementation improves the performance over our non-parallel baselines and how the different memory management techniques affect the overall speedup. Finally, we compare \pname{} with three of the best state-of-the-art BDD libraries, \ie{}, CUDD~\cite{somenzi2012cudd}, Sylvan~\cite{dijk2015sylvan}, and BuDDy~\cite{lind1999buddy}, demonstrating a remarkable speedup boost. The experiments demonstrate the goodness of the proposed package, but given space constraints, extensive benchmarking will be the subject of future work.

In summary, the contributions of \pname{}\footnote{\url{https://luigicapogrosso.github.io/HermesBDD}} are:
\begin{itemize}
\item A computationally faster BDD package, by exploiting multi-threading for parallel processing and concurrent access to a BDD;
\item Multi-platform compatibility (Windows, Linux, and macOS) to accommodate integration within tools from different environments;
\item Support for a static and dynamic memory management mechanism, the possibility to exploit lock-free hash tables, and a novel parallel implementation of the \ite{} procedure based on a higher-level wrapper for threads and futures;
\item High usability and robustness by design, thanks to a development based on engineering principles such as code coverage and continuous integration, along with independence from external software to offer high usability, reliability, and easy maintenance over time.
\end{itemize}

\section{Related Work}
\label{cha:related}

In this section, we provide an overview of the most widely used BDD libraries. For a survey on early packages see~\cite{janssen2003consumer}.

CUDD~\cite{somenzi2012cudd} stands for Colorado University Decision Diagram. It is a single-core package for the manipulation of BDDs, algebraic decision diagrams (ADDs), and Zero-suppressed binary decision diagrams (ZDDs) written in C, with a C++ wrapper.

BuDDy~\cite{lind1999buddy} is a BDD single-core library written in C, with many highly efficient vectorized BDD operations, dynamic variable reordering, automated garbage collection, a C++ interface with automatic reference counting, and much more.

Biddy~\cite{meolic2012biddy} is a BDD package under GPL license, %developed at the University of Maribor. 
whose most distinguishing features are its specially designed C interface and a novel implementation of automatic garbage collection.

CacBDD~\cite{lv2013cacbdd} is a single-core C++ BDD package that implements dynamic cache management, which takes into account the hit rate of the computed table and the available memory. 

BeeDeeDee~\cite{lovato2014thread} is a thread-safe Java library for BDD manipulation. BeeDeeDee allows clients to share a single factory of BDDs, in real parallelism and to reduce the memory footprint of their overall execution at a very low synchronization cost.

Sylvan~\cite{dijk2015sylvan} is a parallel BDD library written in C that provides scalable parallel execution of the standard BDD operations. It supports custom decision diagram terminal types, and it also implements operations on a specialized list of decision diagrams for model-checking.

DecisionDiagrams is a single-core implementation for numerous variants of BDDs that is used at Microsoft Research. Written in C\#, currently, it maintains 100\% code coverage. The library is based on a cache-optimized implementation of decision diagrams~\cite{janssen2001design}.

Based on the work of Lars Arge~\cite{arge1995complexity}, Adiar~\cite{solvsten2022adiar} is a single-core BDD package that makes use of time-forward processing to improve the I/O complexity of BDD manipulation. This achieves efficient manipulation of BDDs, even when they outgrow the memory limit of a given machine.

In~\cite{miyasaka2019simple}, Miyasaka \etal{} describe a simple BDD package without dynamic variable reordering, which is much faster than a conventional BDD package with reordering. The proposed BDD package is used in logic optimization with permissible functions. Moreover, in \cite{miyasaka2020iwls}, Miyasaka presents a framework to compare BDD packages through auto-tuning.

\section{Methodology}
\label{cha:method}

In this section, we present the algorithms and the techniques developed for our efficient parallelization. Due to space constraints, for a technical glance and interesting properties of BDDs, refer to~\cite{akers1978binary,bryant1986graph}.

%%%%%%%%%%%%%%%%%%%%%%%%%
\subsection{The Multi-Core \ite{} Algorithm}
\label{subsec:pite}

\begin{algorithm}[t]
\begin{small}
\caption{$\textsc{p\_ite()}$. The multi-core \ite{}.}\label{alg:para_ite}
$x\leftarrow{}\textsc{top}(f,g,h)$\;
$future\_t\leftarrow{}\textsc{async}(\textsc{ITE},f_x,g_x,h_x)$\;
$t\leftarrow{}future\_t.\textsc{Await.result()}$\;
$future\_e\leftarrow{}\textsc{async}(\textsc{ITE},f_{x'},g_{x'},h_{x'})$\;
$e\leftarrow{}future\_e.\textsc{Await.result()}$\;
\end{small}
\vspace{+2em}
\end{algorithm}
In \pname{} we parallelize the \ite{} function treating the two recursive calls as independent tasks. Starting from the task-based parallel flow of~\cite{van2013multi}, we rewrote it with C++ primitives without using external libraries, which allowed us to introduce a more efficient hash table management mechanism (as we will see in Sec.~\ref{subsec:comparison}). In particular, this flow gives us a simple way to use multiple threads through primitives: instead of making a recursive call to execute the \ite{} function, start a thread at each recursive step, then wait for the thread to finish. With this implementation, the only synchronization between workers is when the results of suboperations are stored in the unique hash table. This table is shared globally, in order to prevent workers from computing a suboperation that was finished already by some other worker. This technique seems to be best suited to parallelize BDDs. We have tried and compared other implementations, \ie{}, parallelism on different levels of the tree. But, given the nature of BDDs, these techniques present an important task overhead, vs. a negligible, or often negative, speedup. Alg.~\ref{alg:para_ite} shows the pseudocode of our implementation. Furthermore, \pname{} provides an option to the user, at compile-time, to decide whether to use the multi-core or sequential implementation of the \ite{} algorithm.

To this end, we use the C++ \verb!std::async()! function, which is a high-level wrapper for threads and futures, followed by the matching function to retrieve the results of the computation. Standard C++ provides \verb!std::thread()! which is a fairly low-level construct, and so its usage is often more cumbersome and error-prone than desired. Instead, \verb!std::async()! automatically creates a thread to call the thread function and it conveniently returns an object \verb!std::future! without the hurdle of manual thread management and decoupling the task from the result.

\begin{algorithm}[t]
\begin{small}
\caption{$\textsc{update\_utable()}$. Creates a BDD node using the unique hash table to ensure that there are no duplicates.}\label{alg:make}
\KwData{$(x,t,e)$}
\If{$t=e$}{
return $e$\;}
\eIf{$\textsc{is\_complemented}(e)$}{
$v\leftarrow{}\textsc{node}(x,t,e)$\;
$n\leftarrow{}\textsc{lookup\_or\_create}(v)$\;
return $\textsc{complement}(n)$\;}
{
$v\leftarrow{}\textsc{node}(x,t,e)$\;
return $\textsc{lookup\_or\_create}(v)$\;}
\end{small}
\vspace{+2em}
\end{algorithm}

To ensure that the results are canonical reduced BDDs, we use the $\textsc{update\_utable}()$ method, as shown in Alg.~\ref{alg:make}. Specifically, the $\textsc{lookup\_or\_create}()$ function checks atomically if data is already in the unique hash table, and if not, it adds it. Finally, in order to enforce canonicity (\ie{}, complement only on 1 edge), we use the functions $\textsc{is\_complemented}()$ and $\textsc{complement}()$.

%%%%%%%%%%%%%%%%%%%%%%%%%
\subsection{A Lock-Free Unique Hash Table for Multi-Core BDDs}

Traditionally, concurrency issues, such as data race, are solved by locks, providing mutual exclusion. Since blocked processes must wait, locks have a negative impact on the speedup of parallel programs. A lot of literature has been dedicated to developing non-blocking algorithms, specifically, Herlihy \etal{}, in~\cite{herlihy2020art}, distinguish between \textit{lock-free}, \textit{wait-free}, and \textit{lock-less} algorithms. %Our implementation falls in the first category.

In particular, we use a lock-free unique hash table, which is implemented using the \verb!std::atomic_flag!. Based on this class, we build a spinlock in order to protect the critical section. Specifically, a spinlock is a lock that causes a thread trying to acquire it to wait in a loop while repeatedly checking whether the lock is available. The use of spinlocks is particularly recommended when the critical section is supposed to perform a minimal amount of work, \ie{}, the spinlock is held for a very short period of time, as in our case. Also, it operates faster compared to mutex since context switching is reduced. Spinlock does not cause the thread to be preempted but instead, it keeps on spinning until the lock on the resource is released.

\begin{algorithm}[t]
\begin{small}
\caption{$\textsc{lookup\_or\_create()}$. Insert an entry into the unique hash table.}\label{alg:insert}
\KwData{$d$}
$hash\leftarrow{}\textsc{calculate\_hash}(d)$\;
$index\leftarrow{}hash\;\%\;elems$\;
$table\_slot\leftarrow{}table[index]$\;
$\textsc{spinlock\_protector}(table\_slot)$\;
$\textsc{insert\_in\_utable}(d)$
\end{small}
\end{algorithm}
The pseudocode for inserting a new value in the unique hash table is given in Alg.~\ref{alg:insert}. This computes the hash, calculates the index, acquires the lock, and finally writes the data into the table. The $\textsc{get\_from\_utable()}$ algorithm works in exactly the same way. This is not reported, since it differs from Alg.~\ref{alg:insert} only by line 5, where it calls the function that compares the parameters and returns the result value.

%%%%%%%%%%%%%%%%%%%%%%%%%
\subsection{The Memory Management Mechanism}
\label{subsec:mm}
BDD algorithms are considered memory intensive since they have little computation for each unit of memory access. Hence, memory allocation techniques for the tables play an important role, and have a great effect, on the performance of the implementations of a BDD package. In \pname{}, we implemented both dynamic and static memory allocation techniques in order to exploit fine-grain parallelism. Also in this case, at compile-time, \pname{} provides an option for the users in order to select the dynamic or the static memory allocation mechanism.

As dynamic memory management, we implemented a simple but effective technique based on doubling the memory space required by the tables. At the beginning of the process, $M$ bytes of memory are allocated to store $N$ nodes. If this space is not enough during program execution, more space of size $M*2$ will be allocated. Since the table is shared by all nodes created by the library, this allows reusing memory. For simplicity and efficiency, there is no strategy for cleaning up the slot table after a given node is removed. % As we will show, in this memory allocation scheme the execution is slightly slower than in the static case, due to the overhead of addressing the correct block of dynamically allocated memory.

In the static allocation technique, instead, a contiguous slice of $M$ bytes of memory is reserved at the start of the process. In this case, the variables get allocated permanently, until the program executes or the function call finishes, and once the memory is allocated, the memory size cannot change, so it is more efficient than dynamic allocation.

\section{Experiments}
\label{cha:experiments}

In this section, we perform quantitative and qualitative analyses to demonstrate the potentiality of our \pname{} package. Experiments were carried out on a 64-bit 32-core AMD Ryzen Threadripper 1950X CPU 3.4GHz machine. The library is tested for compilation using GCC (minimum required: 10.2), Clang (minimum required: 11.0), and MSVC (minimum required: 19.20).

We ran benchmarks from the \nqueens{} problem, using a simple SAT solver based on~\cite{bright2020effective}. In particular, we would like to emphasize that further experiments, as well as systematic comparisons with different parallel tools, will be the subject of future work. Here, also due to space constraints, we focus only on a single well-known problem; nonetheless, we aim to show the potentialities of \pname{} with no interest in presenting a benchmarking paper on BDDs.

%%%%%%%%%%%%%%%%%%%%%%%%%
\subsection{The Speedup Latency \wrt{} our Baselines}

\begin{table}[t]
\centering
\caption{\pname{} non-parallel execution time based on the \nqueens{} problem complexity. Values on the average of 50 samples using the static memory allocation on a 32-core machine.}
\label{tab:nqueens_baseline}
\begin{tabular}{cccc}
\toprule 
\textbf{} & \textbf{$\boldsymbol{6\times{}6}$} & \textbf{$\boldsymbol{7\times{}7}$} & \textbf{$\boldsymbol{8\times{}8}$} \\
\midrule
\textbf{\pname{} baseline} & $15.85$ & $79.99$ & $423.33$ \\
\bottomrule
\end{tabular}
\end{table}

\begin{table}[t]
\centering
\caption{\pname{} parallel speedup based on the \nqueens{} problem complexity and the n. of cores. Speedup was obtained from the average of 50 samples using the static memory allocation.}
\label{tab:nqueens_speedup}
\begin{tabular}{ccccc}
\toprule 
\textbf{Chessboard} & \textbf{2 Core} & \textbf{8 Core} & \textbf{16 Core} & \textbf{32 Core} \\
\midrule
$\boldsymbol{6\times{}6}$ & $1.59\!\times{}\!$ & $2.00\!\times{}\!$ & $2.38\!\times{}\!$ & $2.95\!\times{}\!$ \\
$\boldsymbol{7\times{}7}$ & $2.08\!\times{}\!$ & $2.75\!\times{}\!$ & $3.96\!\times{}\!$ & $4.69\!\times{}\!$ \\
$\boldsymbol{8\times{}8}$ & $4.56\!\times{}\!$ & $9.25\!\times{}\!$ & $10.69\!\times{}\!$ & $\boldsymbol{\underline{18.73\!\times{}\!}}$ \\
\bottomrule
\end{tabular}
\end{table}
Tab.~\ref{tab:nqueens_baseline} shows the result on an average of 50 samples of our baselines on the \nqueens{} problems with the $6\times{}6$, $7\times{}7$, and $8\times{}8$ chessboards, using the static memory allocation mechanism. 

Tab.~\ref{tab:nqueens_speedup} shows the result of our multi-core implementation in terms of speedup, where the speedup latency is computed as $S=T_{ms}(no\_parallel)/T_{ms}(parallel)$, on the same chessboards, using the same number of cores.

Specifically, comparing Tab.~\ref{tab:nqueens_baseline} with Tab.~\ref{tab:nqueens_speedup}, several facts emerge: \textit{i)} smaller models (\eg{}, $6\times{}6$ and $7\times{}7$ chessboard) have lower speedups \wrt{} larger models, which exhibit the best speedups. \textit{ii)} This implies that the speedup is increasing with the size of the Boolean formula. \textit{iii)} Finally, our implementation scales well with the number of cores as long as the problem is sufficiently complex, yielding a speedup of up to $18.73\!\times{}\!$ in the $8\times{}8$ case.

%%%%%%%%%%%%%%%%%%%%%%%%%
\subsection{Comparison \wrt{} other BDD Packages}
\label{subsec:comparison}

\begin{table}[t]
\centering
\caption{\pname{} \wrt{} CUDD~\cite{somenzi2012cudd}, Sylvan~\cite{dijk2015sylvan} and BuDDy~\cite{lind1999buddy} execution time and memory space required. Results were obtained on average from 50 samples of the \nqueens{} problem, using a 32-core machine, and the static memory allocation.}
\label{tab:hermesbdd_wrt_others}
\begin{tabular}{ccccc}
\toprule 
\textbf{} & \textbf{\pname{}} & \textbf{CUDD} & \textbf{Sylvan} & \textbf{BuDDy} \\
\midrule
\textbf{Time (ms)}   & $\boldsymbol{\underline{23.33}}$ & $26.90$ & $33.75$ & $46.50$ \\
\textbf{Memory (GB)} & $\boldsymbol{\underline{0.5}}$   & $1.3$   & $1.9$   & $0.7$ \\
\bottomrule
\end{tabular}
\end{table}
As a comparative approach, we consider CUDD, Sylvan, and BuDDy, which are three of the most important state-of-the-art BDD packages. Tab.~\ref{tab:hermesbdd_wrt_others} reports the results on an average of 50 samples of the \nqueens{} problem with an $8\times{}8$ chessboard, using a 32-core machine, and the static memory allocation mechanism. Specifically, we decided to use static memory allocation for this experiment because it is the memory management technique that allows us to push performance to the maximum in terms of computation time.

In summary, as we can see from Tab.~\ref{tab:hermesbdd_wrt_others}, the combination of using a lightweight approach for parallelism, a lock-free hash table technique, and a static memory allocation yields better performances both for execution time and memory space.

%%%%%%%%%%%%%%%%%%%%%%%%%
\subsection{The Impact of Memory Allocation}
\begin{figure}[t]
\begin{center}
\includegraphics[width=0.90\linewidth]{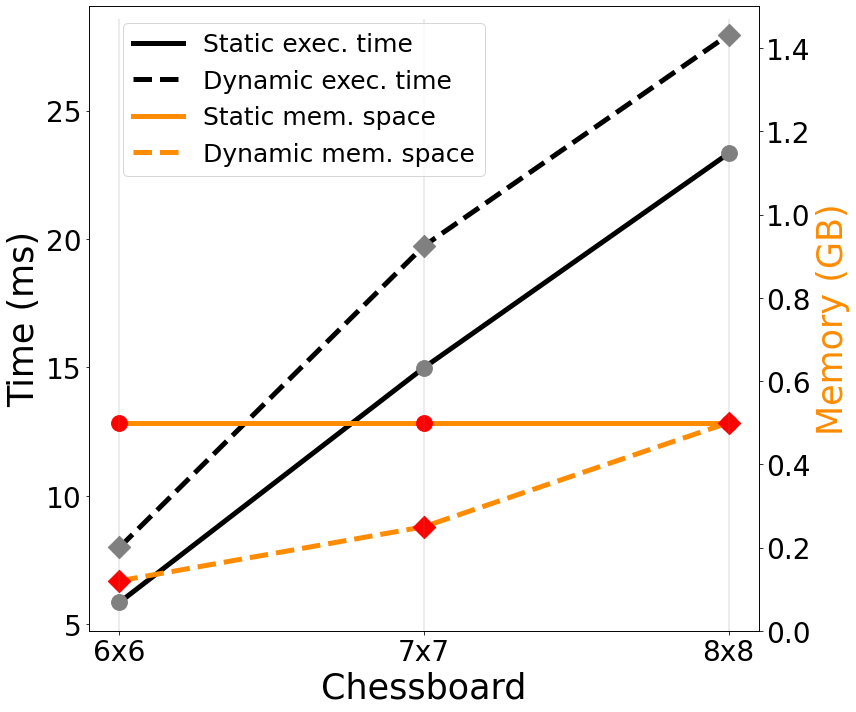}
\end{center}
\caption{\pname{} results on the impact of memory allocation in terms of time and memory space, on an average of 50 samples of the \nqueens{} problem on a 32-core machine.}
\label{fig:hermesbdd_mm}
\end{figure}
In this experiment, we evaluate the impact of both static and dynamic memory allocation techniques in terms of execution time and memory space. Specifically, Fig.~\ref{fig:hermesbdd_mm} shows the results of a single test in which we run, sequentially, the \nqueens{} problem on a $6\times{}6$, then on a $7\times{}7$, and finally on an $8\times{}8$ chessboard on an average of 50 samples of the \nqueens{} problem on a 32-core machine. As mentioned in Sec.~\ref{subsec:pite}, thanks to the unique hash table the $7\times{}7$ case benefits from the $6\times{}6$ storage, just as the $8\times{}8$ case benefits from the $7\times{}7$ storage. 

The behaviors explained in Sec.~\ref{subsec:mm} are confirmed: a dynamic memory allocation guarantees a more efficient memory occupancy, but the overall execution time is slower than static memory allocation due to the presence of overhead caused by the management of the memory allocation at run time (\eg{}, the increase of execution time in percentage is $36\%$ for the $6\times{}6$ chessboard, down to $19\%$ for the $8\times{}8$ chessboard). On the other hand, static memory allocation provides better performance in terms of execution time but at the expense of inefficient memory management (\eg{}, in the case of the $6\times{}6$ chessboard, more MBs are allocated than required).

\section{Conclusions}
\label{cha:conclusions}

In this paper, we presented \pname{}, a multi-core and multi-platform package for BDD manipulation. We designed and implemented three different algorithms to support a parallel implementation of BDD operations: a multi-core \ite{} procedure based on a higher-level wrapper for threads and futures, a lock-free hash table, and both a static and dynamic memory allocation mechanism.

The performance demonstrated a significant speedup. Thus, we can say that the experiments validate the proposed package, whereas more extensive benchmarking will be the subject of future work.

%\section*{Acknowledgements}
%The work has been partially supported by the Italian Ministry of Education, University and Research (MIUR) with the grant ``Dipartimenti di Eccellenza'' 2018-2022.

%%%%%%%%% BIBLIOGRAPHY.
\bibliographystyle{IEEEtran}
\bibliography{bibi}

%%%%%%%%%%%%%%%%%%%%%%%%%
\newpage
\appendix

%%%%%%%%%%%%%%%%%%%%%%%%%
\subsection{Introduction}
BDDs enabled breakthroughs in the area of formal verification, by advancing the capability of state-of-the-art \textit{model-checking} by several orders of magnitude in terms of the size of state spaces that can be successfully explored. While BDDs have been applied successfully to discrete systems~\cite{Burch1992}, exploiting them to compute the evolution of cyber-physical systems~\cite{hscc2020} allows introducing an efficient discretized representation of states even for non-linear dynamical systems~\cite{arch2019nln}. So BDDs and SAT are required engines in the CAD toolbox~\cite{intro-procieee2015}. Finally, another area in which these diagrams are particularly useful is that of \textit{test generation}, \ie{}, finding a set of inputs that can be used to confirm that a given implementation performs correctly~\cite{paul2006finding,ahishakiye2021mc}.

Model-checking (or property-checking) is the method that creates abstractions of complex systems to verify if their function meets a given specification. Algorithms for model-checking were initially implemented in an explicit-state manner~\cite{clarke1986automatic}. This meant that all automata involved in verification were represented listing their states explicitly. From a theoretical perspective, the data structure used to represent the automata makes no difference, but from a practical perspective, it means that model checkers could only handle automata with few reachable states~\cite{chaki2018bdd}. So, practical hardware verification via model-checking called for a computational breakthrough, which appeared in the form of BDD-based symbolic model-checking~\cite{burch1992symbolic}, \ie{}, representing all states using functions, instead of storing them individually. This fact enabled practical verification of industrial systems, beginning with hardware~\cite{burch1994symbolic}, and extending to software~\cite{ball2001bebop}. Moreover, it led to important developments for decision diagrams, such as new types of BDDs (\eg{},~\cite{minato2001zero}) and new variable-ordering heuristics (\eg{},~\cite{aziz1994bdd}). For example, QMDDs~\cite{miller2006qmdd} are a kind of BDDs specifically designed to represent and manipulate matrices that allow compact simulation and specification of reversible quantum gates and circuits.

Although direct exploration of parameters is possible in some cases, it is also always computationally intensive~\cite{bryant1995binary}. Therefore, techniques to increase the performance of model-checking tools are under constant development. Until the last decade, the usual policy to improve the memory and computation performance was to increase the CPU frequencies and implement various hardware and software optimizations~\cite{chan2001optimizing}. However, with recent developments in parallel hardware architectures, the challenge to push the frontiers of BDD applicability calls for the deployment of BDD packages on multi-processor architectures. This can be accomplished using various techniques, such as better algorithms~\cite{cormen2022introduction}, and parallel computing using multiple processors~\cite{rauber2013parallel}. In this sense, having a BDD package that can perform optimized operations through efficient algorithms, and perform them on multiple processors, will enable us to push the frontiers of formal verification with BDDs.

%%%%%%%%%%%%%%%%%%%%%%%%%
\subsection{Background}
In this section, we briefly review the background, definition, and interesting properties of BDDs.

%%%%%%%%%%%%
\subsubsection{Terminology}
Let $\textbf{x}$ denote a vector of Boolean variables $x_1,x_2,\dots{}x_n$, and let $\textbf{a}$ denote a vector of values $a_1,a_2,\dots{}a_n$, $a_i\in{}\{0,1\}$. Let $\textbf{1}$ denote the function that always yields 1, and $\textbf{0}$ the function that always yields 0.

We can define Boolean operations $\land{},\lor{},\oplus{}$, and $\neg{}$ over functions according to the Boolean operations on the underlying elements. So, for example, $f\land{}g$ is a function $h$ such that $h(\textbf{a})=f(\textbf{a})\land{}g(\textbf{a})$ for all $\textbf{a}$~\cite{bryant1992symbolic}.

For function $f$, variable $x_i$ and a binary value $b\in{}\{0,1\}$, we can define a \textit{restriction} of $f$ as the function resulting when $x_i$ is set to value $b$:
\begin{equation}
f|_{x_i\leftarrow{}b}(\textbf{a})=f(a_1,\dots{},a_{i-1},b,a_{i+1},\dots{},a_n)
\end{equation}
In particular, the two restrictions of a function $f$ \wrt{} a variable $x_i$ referred to as \textit{cofactors} of $f$ \wrt{} $x_i$~\cite{brayton1984logic}.

Given the two cofactors of $f$ \wrt{} to variable $x_i$, the function can be reconstructed as:
\begin{equation}
f=(x_i\land{}f|_{x_i\leftarrow{}1})\lor{}(\neg{}x_i\land{}f|_{x_i\leftarrow{}0})
\label{eq:shannon_decomposition}
\end{equation}
This identity is commonly referred to as \textit{Shannon decomposition} of $f$ \wrt{} $x_i$~\cite{brown2003boolean}. 

%%%%%%%%%%%%
\subsubsection{Definition}
Any Boolean function $f:B^n\rightarrow{}B$, where $n$ is the number of variables, and $B\in{}\{0,1\}$, is a mapping from a $n$-bit binary vector, to a single bit output~\cite{elbayoumi2013novel}. This function can be represented as a binary decision diagram, which is a directed acyclic graph (DAG) expressing the Shannon decomposition (Eq.~\ref{eq:shannon_decomposition}) of a Boolean function.

In particular, BDDs are defined as tuples $[V,h,l,v]$ where $V$ represents the set of internal vertices, $h,l:V\rightarrow{}V\cup{}\{0,1\}$ are functions representing the high and low edges of a node, $v$ indicates the variable associated to a vertex, and leaves can be 0 or 1.

A BDD is called \textit{ordered} (OBDD) if each variable is encountered at most once along each path and variables appear in the same order on all paths. So, on every path from the root to a leaf, every variable is encountered at most once.

A BDD is called \textit{reduced ordered} (ROBDD) if it contains no nodes with two identical child nodes and no duplicated sub-graphs. The advantage of a ROBDD is that it is canonical (\ie{}, unique) for a particular function and variable order~\cite{bryant1986graph}. Any BDD can be reduced by applying the following two rules:
\begin{enumerate}
\item Eliminate redundant nodes.
\item Eliminate duplicated sub-graphs by sharing sub-graphs.
\end{enumerate}
This property makes it useful in functional equivalence checking and other operations like functional technology mapping. A path from the root node to the 1-terminal represents a (possibly partial) variable assignment for which the represented Boolean function is true. As the path descends to a low (or high) child from a node, then that node's variable is assigned to 0 (or 1).

%%%%%%%%%%%%
\subsubsection{Complemented Edges}
To decrease memory usage and calculation time, sub-graphs can be reused to represent related Boolean functions by adding attributes to edges in a BDD. Several methods are mentioned in the literature, such as~\cite{miller1997negation,miller1998dual}. 

Introduced by Minato~\cite{minato1990shared} and Brace~\cite{brace1990efficient}, complemented edges (also called negated edges), indicate that leaves 1 and 0 will be switched, negating the Boolean function represented by the node. If an edge is complemented, then it refers to the negation of the Boolean function that corresponds to the node to which the edge points (the Boolean function represented by the BDD rooted in the node). It allows performing complementation in constant time. For example, a ROBDD can be represented more compactly using complemented edges~\cite{somenzi1999binary}. In order to guarantee the canonicity of BDDs with complemented edges, Minato~\cite{minato1990shared} proposed the following constraints:
\begin{itemize}
\item Only 0 is used as the value of a leaf.
\item No complemented edges are allowed on 0-edges.
\end{itemize}

%%%%%%%%%%%%
\subsubsection{The \ite{} Operator}
\begin{table}[t]
\centering
\caption{Implementation of Boolean functions in terms of \ite{} operators.}
\label{tab:ite}
\begin{minipage}{.51\linewidth}
\begin{tabular}{|c|c|c|}
\toprule 
\textbf{No.}   & \textbf{Boolean}   & \textbf{ITE Op.}\\
\midrule
0  & $0$                    & $0$\\
1  & $f\wedge{}g$           & $ITE(f,g,0)$\\
2  & $f\nRightarrow{}g$     & $ITE(f,g',0)$\\
3  & $f$                    & $f$\\ 
4  & $f\nLeftarrow{}g$      & $ITE(f,0,g)$\\
5  & $g$                    & $g$\\
6  & $f\oplus{}g$           & $ITE(f,g',g)$\\
7  & $f\vee{}g$             & $ITE(f,1,g)$\\
\bottomrule
\end{tabular}
\end{minipage}%
\begin{minipage}{.51\linewidth}
\begin{tabular}{|c|c|c|}
\toprule 
\textbf{No.}   & \textbf{Boolean}   & \textbf{ITE Op.}\\
\midrule
8  & $f\mp{}g$              & $ITE(f,0,g')$\\
9  & $f\Leftrightarrow{}g$  & $ITE(f,g,g')$\\
10 & $\overline{g}$         & $ITE(g,0,1)$\\
11 & $f\Leftarrow{}g$       & $ITE(f,1,g')$\\
12 & $\overline{f}$         & $ITE(f,0,1)$\\
13 & $f\Rightarrow{}g$      & $ITE(f,g,1)$\\
14 & $f\barwedge{}g$        & $ITE(f,g',0)$\\
15 & $1$                    & $1$\\
\bottomrule
\end{tabular}
\end{minipage}%
\end{table}
We exploit the observation that all Boolean operators with two arguments can be expressed using the \ite{} operator, with arguments the two Boolean functions or their negations, or the constants 0 and 1.
The \ite{} (\verb!ITE!) is defined as follows:
\begin{equation}
\verb!ITE!(f,g,h)=(f\land{}g)\lor{}(f'\land{}h)
\label{eq:ite}    
\end{equation}
where $f$, $g$ and $h$ are all Boolean functions. The \verb!ITE! can be recursively computed as follows:
\begin{equation}
\verb!ITE!(f,g,h)=\verb!ITE!(v,\verb!ITE!(f_v,g_v,h_v),\verb!ITE!(f_{v'},g_{v'},h_{v'}))
\label{eq:ite_rec}    
\end{equation}
where $v$ is a Boolean variable on which the functions depend, $f_v$ and $f_{v'}$ are the positive and negative cofactors for $f$, respectively (same holds for $g$ and $h$). Such a recursive formulation also allows for efficient storing and indexing of BDDs. In fact, it forms the basic structure of the unique table, in which each entry stores a triplet (variable, right-pointer, left-pointer)~\cite{stornetta1996implementation}.

We define the \ite{} normal form as a Boolean expression using only the \ite{} operator and the constants 0 and 1. In Tab. \ref{tab:ite} we report the equivalent \ite{} operator for each of the 16 possible Boolean operators on two variables.

\RestyleAlgo{ruled}
\begin{algorithm}[t]
\begin{small}
\caption{$\textsc{ite()}$. The \ite{}.}\label{alg:seq_ite}
\KwData{$(f,g,h)$}
\KwResult{$r$}
\eIf{terminal\_case}{
return $computed\_result$\;}
{
\eIf{$\{(f,g,h),r\}$ in cache}{
$r\leftarrow{}\textsc{get\_from\_cache}(\{(f,g,h),r\})$\;
return $r$\;}
{
$x\leftarrow{}\textsc{top}(f,g,h)$\;
$t\leftarrow{}\textsc{ite}(f_x,g_x,h_x)$\;
$e\leftarrow{}\textsc{ite}(f_{x'},g_{x'},h_{x'})$\;
\If{$t=e$}{
return $t$\;}
$r\leftarrow{}\textsc{update\_utable}(x,t,e)$\;
$\textsc{update\_cache}(\{(f,g,h),r\})$\;
return $r$\;}
}
\end{small}
\end{algorithm}
Alg.~\ref{alg:seq_ite} presents the pseudocode of our sequential \ite{} implementation, based on~\cite{brace1990efficient}. The function takes three arguments, which are the pointers to BDDs representing the \textsc{if}, \textsc{then}, and \textsc{else} Boolean functions, and returns a pointer to the resulting BDD.

The algorithm starts by checking if a terminal case applies, where the result can be easily computed. If so, the result is returned. Otherwise, the algorithm checks in the computed table if the result has already been computed, and if so, it returns it. If not, the variable of the root node is determined through the $\textsc{top()}$ function, and the cofactors of the original \ite{} operator are recursively computed. If they are equal, one of them is returned as the result. Otherwise, the function $\textsc{update\_utable()}$ is used to check if the new BDD node has already been created. If so, a pointer to it is returned. If not, the BDD node is first created and then returned, and the information for it is inserted in the computed table to avoid future recomputations.

%%%%%%%%%%%%
\subsubsection{Efficiency Improvements}
BDDs are typically irregular and require an unpredictable number of memory accesses with high demands on memory latency~\cite{yang1998performance}. Researchers have spent much time and effort examining how they can represent these structures more efficiently.

Bryant \etal{} in~\cite{brace1990efficient} described various efficiency improvements that are now standard in a BDD implementation. They include complemented edges, hashing, caching, and garbage collection.

Moreover, the size of BDDs depends heavily on the chosen variable ordering. We suggest referring to~\cite{meinel1998algorithms} by Meinel \etal{} for a thorough discussion on variable ordering.

An approach to BDD implementation that discards the use of pointers in favor of simple arrays is discussed in G. Jansen’s paper~\cite{janssen2001design}. This approach is shown to outperform a pointer-based approach in memory and CPU efficiency, at the expense of more complicated garbage collection.

%%%%%%%%%%%%%%%%%%%%%%%%%
\subsection{The \nqueens{} Problem}
The eight-queens puzzle is the problem of placing eight chess queens mutually non-attacking on an $8\times{}8$ chessboard~\cite{rivin1994n}. Thus, a solution requires that no two queens share the same row, column, or diagonal. In particular, this problem can be posed in more general terms through the following question: \textit{Given $n$, in how many ways can $n$ queens be placed on an $n\times{}n$ chessboard without threatening each other?}

Inspired by~\cite{kunkle2010parallel}, we implement a solution algorithm by constructing a BDD row-by-row that represents whether the row is in a legal state. Then we count the number of satisfying assignments of the accumulation BDD.

%%%%%%%%%%%%%%%%%%%%%%%%%
\subsection{Comparison \wrt{} other BDD Packages}
There are several differences between the implementation in CUDD, Sylvan, and BuDDy, and the one in \pname{} that makes it difficult to compare the performances. For example, CUDD, Sylvan, and \pname{} use \textit{reference counting} for garbage collection, while BuDDy uses the \textit{mark-and-sweep} technique, which requires less accounting work. However, the preallocated tables were large enough that garbage collection did not occur in any of the packages. BuDDy also uses several other optimizations, such as increased memory locality by storing related BDD nodes near each other in the hash table, while Sylvan and \pname{} store BDD nodes at the same position as the hash in the hash table. Finally, BuDDy and CUDD are not thread-safe, as opposed to Sylvan and \pname{}. Finally, we can say that our implementation has an advantage compared to Sylvan's because it features a native implementation of parallelism. Indeed, C++ primitives guarantee complete control of all task-based aspects, whereas Sylvan relying on an external library to handle parallelism does not have these guarantees. This allows us to implement a lock-free mechanism for the management of the hash table, which is more efficient than the lock-less type mechanism used by Sylvan. Moreover, interested readers can refer to~\cite{janssen2003consumer} for a comprehensive discussion on the differences between the implementation of CUDD and BuDDy.

%%%%%%%%%%%%%%%%%%%%%%%%%
\subsection{Conclusions}
The performance demonstrated a significant speedup, such as $18.73\!\times{}\!$ over our non-parallel baselines, and a remarkable performance boost \wrt{} other state-of-the-art BDD packages, such as $1.41\!\times{}\!$ \wrt{} Sylvan. Thus, we can say that the experiments validate the proposed package, whereas more extensive benchmarking will be the subject of future work. Additionally, to fill a gap in the literature, \pname{} %has been developed to be multi-platform, 
comes with a well-documented source code, does not rely on external libraries, and has been designed according to software engineering principles such as CI/CD for reliability and easy maintenance over time.

In future work, we will fine-tune the current implementation by adding functions for composition, Boolean quantification, and BDD minimization, exploiting further optimizations. %In addition, we will experiment with other hardware technologies, \eg{}, GPU computing, to make our library more efficient.

\end{document}